\documentclass{article}

\usepackage{arxiv}

\usepackage[utf8]{inputenc} 
\usepackage[T1]{fontenc}    
\usepackage{hyperref}       
\usepackage{url}            
\usepackage{booktabs}       
\usepackage{amsfonts}       
\usepackage{nicefrac}       
\usepackage{microtype}      
\usepackage{lipsum}		
\usepackage{graphicx}
\usepackage{natbib}
\usepackage{doi}
\usepackage{listings}
\usepackage{xcolor}
\usepackage{algorithm}
\usepackage{algpseudocode}

\definecolor{codegreen}{rgb}{0,0.6,0}
\definecolor{codegray}{rgb}{0.5,0.5,0.5}
\definecolor{codepurple}{rgb}{0.58,0,0.82}
\definecolor{backcolour}{rgb}{1,1,1}
\lstdefinestyle{bash}{
    language=bash,
    backgroundcolor=\color{backcolour},
    commentstyle=\color{codegreen},
    keywordstyle=\color{blue},
    numberstyle=\tiny\color{codegray},
    stringstyle=\color{codepurple},
    basicstyle=\footnotesize\ttfamily,
    breakatwhitespace=false,
    breaklines=true,
    captionpos=b,
    keepspaces=true,
    numbers=left,
    numbersep=5pt,
    showspaces=false,
    showstringspaces=false,
    showtabs=false,
    tabsize=2
}
\lstdefinestyle{java}{
    language=Java,
    backgroundcolor=\color{backcolour},   
    commentstyle=\color{codegreen},
    keywordstyle=\color{magenta},
    numberstyle=\tiny\color{codegray},
    stringstyle=\color{codepurple},
    basicstyle=\footnotesize,
    breakatwhitespace=false,         
    breaklines=true,                 
    captionpos=b,                    
    keepspaces=true,                 
    numbers=left,                    
    numbersep=5pt,                  
    showspaces=false,                
    showstringspaces=false,
    showtabs=false,                  
    tabsize=2,
        literate= 
     *{0}{{{\color{magenta}0}}}{1}
      {1}{{{\color{magenta}1}}}{1}
      {2}{{{\color{magenta}2}}}{1}
      {3}{{{\color{magenta}3}}}{1}
      {4}{{{\color{magenta}4}}}{1}
      {5}{{{\color{magenta}5}}}{1}
      {6}{{{\color{magenta}6}}}{1}
      {7}{{{\color{magenta}7}}}{1}
      {8}{{{\color{magenta}8}}}{1}
      {9}{{{\color{magenta}9}}}{1}
      {:}{{{\color{codegreen}{:}}}}{1}
      {,}{{{\color{codegreen}{,}}}}{1}
      {\{}{{{\color{codegreen}{\{}}}}{2}
      {\}}{{{\color{codegreen}{\}}}}}{2}
      {[}{{{\color{codegreen}{[}}}}{3}
      {]}{{{\color{codegreen}{]}}}}{3},
}

\lstdefinestyle{gherkin}{
    language={}, 
    backgroundcolor=\color{white},   
    commentstyle=\color{codegreen},
    keywordstyle=\color{blue},
    numberstyle=\tiny\color{codegray},
    stringstyle=\color{codepurple},
    basicstyle=\footnotesize\ttfamily,
    breakatwhitespace=false,         
    breaklines=true,                 
    captionpos=b,                    
    keepspaces=true,                 
    numbers=left,                    
    numbersep=5pt,                  
    showspaces=false,                
    showstringspaces=false,
    showtabs=false,                  
    tabsize=2,
    morekeywords={Scenario, Given, Then, And, Background, Feature, When} 
}

\title{Advancing BDD Software Testing: Dynamic Scenario
Re-Usability And Step Auto-Complete For Cucumber Framework}

\author{ \href{https://orcid.org/0000-0002-0724-9197}{\includegraphics[scale=0.06]{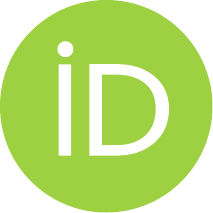}\hspace{1mm}Ali Hassaan Mughal} \\
	Software Developer-II, \\
        Xpressdocs Partners Pvt. Ltd.\\
        MSc. Computer Science,\\
	Kansas State University.\\
}

\renewcommand{\shorttitle}{\textit{arXiv} Template}

\hypersetup{
pdftitle={Advancing BDD Software Testing: Dynamic Scenario
Re-Usability And Step Auto-Complete For Cucumber Framework with Gherkin},
pdfsubject={q-bio.NC, q-bio.QM},
pdfauthor={Ali H.~Mughal},
pdfkeywords={Software Testing, Software Testing Optimization, Automated Software Testing, Cucumber BDD, Gherkin BDD Testing, VSCode Autocompletion, Test Code Generation},
}

\begin{document}
\maketitle

\begin{abstract}
This paper presents and implements the re-usability of scenarios within scenarios for behavior-driven
development (BDD) Gherkin test scripts in the Cucumber Java framework. Though the focus of the presented work is on scenario re-usability
through an implementation within the Cucumber BDD Java framework, the paper also dives a little
into the limitations of Cucumber’s single-threaded scenario execution model. This implementation
increases the modularity and efficiency of the test suite. The paper also discusses VSCode’s
step definition auto-completion integration, simplifying the test script writing process. This functionality
is handy to Quality Assurance(QA) test writers, allowing instant access to
relevant step definitions. In addition, the use of these methods in a popular continuous integration
and delivery platform Jenkins as a Maven Java project is discussed. This integration with Jenkins, facilitates for more
efficient test automation for continuous deployment scenarios. Empirical research and practical
applications reveal significant improvements in the speed and efficiency of test writing, which is
especially valuable for large and complex software projects. Integrating these methods into
traditional sequential BDD practices paves the way towards more effective, efficient, and sustainable test automation
strategies. 
\end{abstract}


\section{Introduction}
\cite{smart2023bddbook, Cucumber_BDD_Christy_Barus_2019} BDD is a very relevant methodology in modern software development and testing, as it bridges the gap between technical implementation and business requirements. Cucumber stands out as a leading tool in this domain, facilitating the definition of application behavior in a human-readable language \cite{geomar59704_cucu_2023, Cucumber_BDD_Christy_Barus_2019}. Despite its widespread adoption, the Cucumber framework faces challenges in scenario maintenance and scalability, particularly when operating within single-threaded execution models. This paper tries to address these challenges by proposing an approach to use scenario re-usability within the Cucumber framework in Java, and talks towards its application to be run on Jenkins, a leading continuous integration and delivery platform, to run the cucumber BDD, to enable calling a scenario inside another scenario; scenario re-usability (SR) functionality.

The main contribution of this paper is the development and workout of a method that mostly reuses the already built functionality that allows it to execute reusable scenario/(s) (RS) within scenario/(s). Keeping the process as a single-threaded, and without calling cucumber as a separate process. Thus, goes around its traditional execution limitations. This method not only addresses efficiency concerns in single-threaded environments but also significantly enhances the modularity and maintainability of BDD test suites. Cucumber supports backgrounds, which are run once before every script, but that does not solve the problem of re-usability within a scenario. With this functionality in place, a QA tester can write reusable scenarios, and have more access and better modularity for the testing feature set.

Listing~\ref{lst:gherkincode} shows an example of Gherkin BDD feature file.
\begin{lstlisting}[style=gherkin, caption=An example of how a gherkin .feature file looks like with sample scenarios for testing user access., label={lst:gherkincode}]
@blog_management @access_control
Feature: Multiple Site Support
  Only blog owners can post to a blog, except administrators, who can post to all blogs.

  Background:
    Given a global administrator named "{Admin_Name}"
    And a blog named "{Admin_Blog}" owned by "{Admin_Name}"
    And a customer named "{User_Name}"
    And a blog named "{User_Blog}" owned by "{User_Name}"

  @user_posting
  Scenario: {User_Name} posts to their own blog
    Given I am logged in as {User_Name}
    When I try to post to "{User_Blog}"
    Then I should see "{Success_Message}"

  @cross_posting_error
  Scenario: {User_Name} tries to post to somebody else's blog, and fails
    Given I am logged in as {User_Name}
    When I try to post to "{Admin_Blog}"
    Then I should see "{Failure_Message}"

  @admin_privileges
  Scenario: {Admin_Name} posts to a client's blog
    Given I am logged in as {Admin_Name}
    When I try to post to "{User_Blog}"
    Then I should see "{Success_Message}"

\end{lstlisting}

In addition to implementing the re-usability, this paper also focuses on the usability of VSCode's step definition auto-complete functionality extension \cite{krechik2017cucumber}, to showcase its potential in easing the scenario writing and management process. This ability to reuse scenarios is especially advantageous for beginner QA test writers, as it provides easy access to and navigation of step definitions, hence reducing the learning curve and improving productivity in test development. Though \cite{intellijidea} also provides similar auto-complete functionality, it is a paid integrated development environment (IDE) software for professional use. 
The paper conveys an idea towards significant improvements in test framework development speed and execution efficiency. These advancements are particularly beneficial for large-scale and complex software projects, where traditional BDD practices may fall short. The integration of these innovative methods with established BDD practices suggests a pathway towards more effective, efficient, and sustainable test automation strategies, especially in environments utilizing Jenkins for continuous integration and delivery. The implementation of this project is done in Java language, though this same seemed achievable in Cucumber for Python and Javascript per the understanding of the author. Still, that is out of the scope of this paper.

The main contributions of the paper are as follows:
\begin{itemize}
    \item  Reusing Gherkin scenarios within scenarios in the Cucumber framework aimed at enhancing the speed, modularity, and maintainability of test writing.
    \item To generate dynamic ENUM types, to improve integration with VSCode auto-complete functionality, thereby enhancing visibility for QA test script writers and reducing the need for manual scenario file searches.
    \item To traverse through and successfully implement BDD scenario re-usability and to be used while running a Maven test project on Jenkins, demonstrating an innovative approach to adapt this functionality to running with Jenkins.
    \item To demonstrate the setup for step definition auto-complete functionality in VSCode \cite{visualstudiocode} with suggestions and its combination with dynamic ENUMs, highlighting its benefits for writing BDD tests and improving the efficiency of test development.
    \item To present a use case of scenario re-usability in a large-scale e-commerce company with expanded customer bases where each feature needs to be tested under different conditions, emphasizing its effectiveness in handling similar functionalities across different contexts.
\end{itemize}
\begin{figure}[htbp]
\centering
\includegraphics[width=\linewidth]{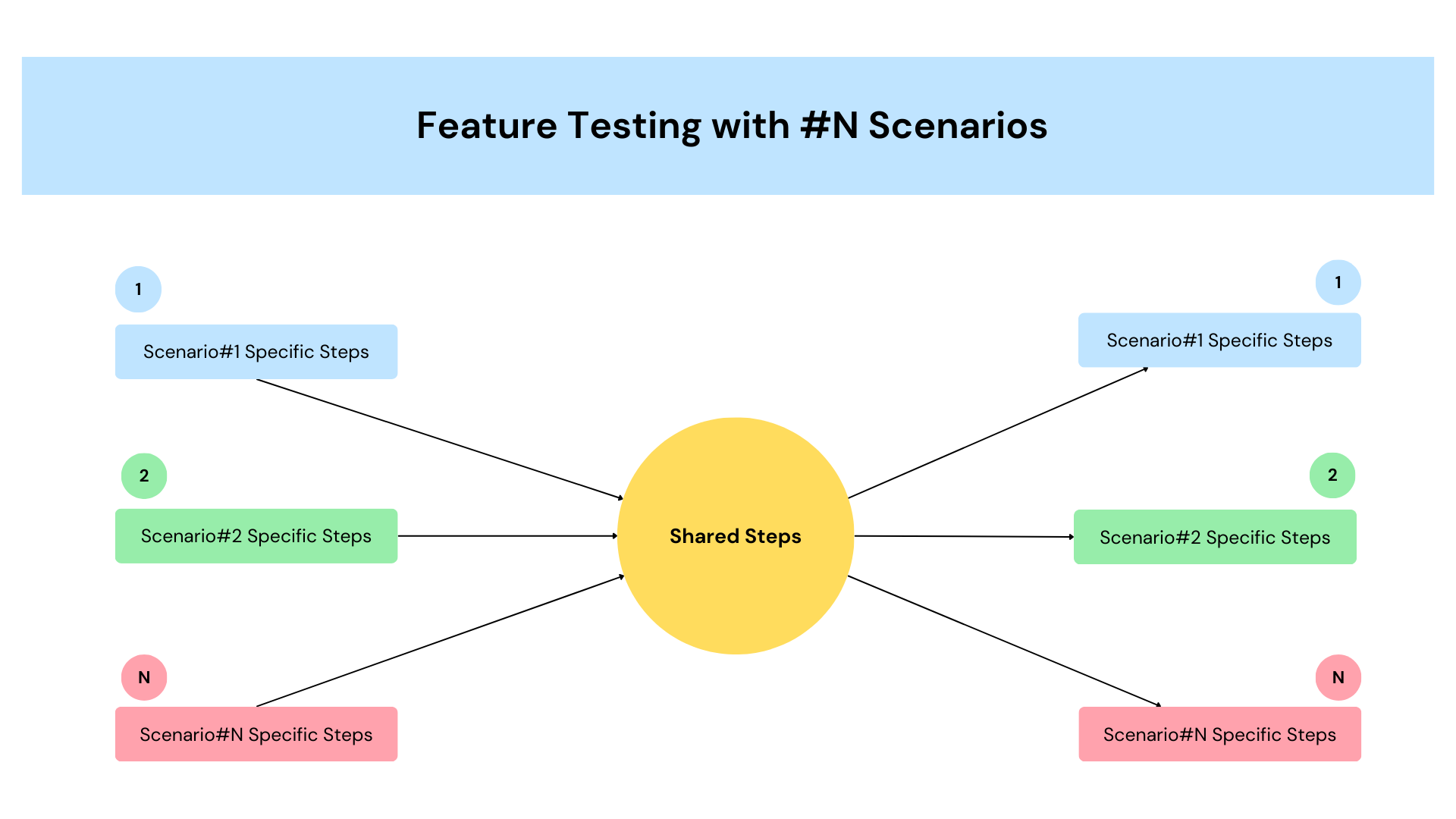}
\caption{Depiction of a Feature being tested with \#N Scenarios using a shared set of steps(Reusable Scenario).}
\label{fig:my_label}
\end{figure}

\section{Advantages of the Proposed Approach}

This paper presents re-usability of scenarios within the Cucumber framework. This improves the efficiency and flexibility of creating and maintaining behavior-driven development (BDD) test cases and helps in streamlining the test writing and hence the testing process. 

\subsection{Better re-usability of tests}
The main advantage of this approach is greater re-usability of test cases. By allowing one scenario to effectively call another, by allowing common test functions to be reused across different test cases. This re-usability is especially useful when testing similar features in different settings or environments.

\subsection{Faster Test Development}
Integrating VSCode step definition auto-completion with this dynamic ENUM types greatly speeds up the test writing process. This setting enables faster test generation, making it especially useful for new QA test writers who now have easy access to relevant step definitions while writing a test case script.

\subsection{Centered Test Case Maintenance}
For scenarios where there occurs User Interface (UI) updates or other changes in the functionality being tested(inside a reusable scenario), only reusable steps need to be updated. This localized update process reduces the time and effort required to maintain test suites, especially in dynamic development environments where frequent changes takeplace. This ability of modularity also lends itself to greater flexibility, allowing for easy adaptation to different testing requirements.

\subsection{Scaling for Large-Scale Projects}
For large-scale and complex software projects, this method offers a scalable solution. It enables efficient management and execution of extensive test suites, ensuring consistent quality and performance across the project lifecycle.

\subsection{Integration with Jenkins}
The paper demonstrates the application of these methodologies within Jenkins, highlighting a seamless integration process. This integration facilitates more efficient test automation in continuous deployment scenarios, a critical aspect for modern software development pipelines.

\section{Literature Review}

In the realm of software development, particularly focusing on testing methodologies, several notable studies and theses offer insights into innovative approaches. 

\cite{litref1} compares Cucumber and Robot framework as BDD automated testing tools, focusing on functionality, reliability, usability, performance efficiency, and portability, recommending Cucumber for novice Quality Assurance Testers (QATs) and Robot framework for more experienced QATs due to its concise syntax and extensive keywords.
\cite{litref2} introduces a method for testing drivers on microcontrollers, using a Device Under Test, a double device, and a computer to emulate real external peripherals, successfully validating the solution with different protocols.
\cite{litref3} identifies key characteristics of Behaviour Driven Development (BDD) through literature analysis and current toolkits, providing a foundational understanding of BDD.
\cite{litref4} reports on the challenges of using BDD in a large public health-related BI project, suggesting that BDD may not be appropriate in the early development stages of BI projects due to the complex business understanding required.
\cite{litref5} discusses challenges in automating the generation of test classes from use cases in narrative form, and analyzing toolkits for integrating BDD libraries like JBehave with development environments.
\cite{litref6} develops a tool for generating BDD test case codes from scenario definitions, facilitating automated testing for entry-level programmers.
\cite{litref7} proposes the GoBeAT methodology for MAS test suite specification in the robot soccer domain, demonstrating its effectiveness in a RoboCup case study.\\\\
\cite{litref8} implements an automated testing framework addressing technical and social aspects, aiming to align client requests with the technical area using BDD.
\cite{litref9} uses Gherkin to capture requirements from legislation in a Gift Aid system development case study, demonstrating the alignment of formal and agile development.
\cite{litref10} introduces a behavior-driven cloud-based framework for agile product line engineering, focusing on automated testing for managing core assets and variations.
\cite{litref11} discusses the application of BDD in large projects, suggesting a taxonomy for its successful application based on various project aspects.
\cite{litref12} demonstrates a tool for documentation by extending the BDD approach to generate readable documents from a JUnit test suite.
\cite{litref13} discusses using Gherkin CNL and MT for global product information development in McAfee, aiming to integrate it into the SDLC.
\cite{litref14} describes an approach for augmenting system verification with automated testing, enhancing the human verification process.
\cite{litref15} adapts BDD to testing ALMA telescope calibration software, focusing on communication between developers, testers, and scientists.
\cite{litref16} proposes an approach to bridge the gap between business specialists and software designers, using a mind model specification to transform scenario definitions into a conceptual model.
\cite{litref17} presents a model-driven approach for behavior-driven GUI testing, transforming BDD-like requirements and GUI descriptions into executable test cases.
\cite{litref18} discusses the use of BDD, particularly Gherkin, and RSpec, in DevOps for dependable Cyber-Physical Systems.\\\\
\cite{litref19} explores behavior-driven graphical user interface testing, introducing a specification language for automatically executable test cases from BDD-like feature descriptions.
\cite{litref20} introduces a behavior-driven Quality First Agile Testing approach for cloud service, focusing on a black box approach to address the quality engineering challenges in cloud services.
\cite{litref21} proposes a Gherkin syntax extension for parameterizing network switch configurations in BDD test specifications, effectively reducing the lines needed for configuration descriptions.
\cite{litref22} discusses behavior-driven model-based re-engineering for cloud testing, emphasizing the design and validation of test instances with GraphWalker models and domain-specific languages.
\cite{litref23} proposes an approach for behavior-driven development of software product lines, focusing on automatically composing and executing test suites according to selected features.
\\\\
\cite{litref24} introduces the Accountability Driven Development framework for ML systems, focusing on recording accountability information and transforming requirements into specific scenarios.
\cite{litref25} proposes ontological syntax highlighting to enhance the readability and understandability of the code.
\cite{litref26} presents a paper-based game for teaching BDD concepts in a practical, competitive, and fun way.
\cite{litref27} discusses combining STPA and BDD for safety analysis and verification in agile development.\\\\
\cite{litref28} introduces SS-BDD, a framework for building and running spreadsheet test scenarios using BDD, achieving high fault-detection effectiveness.
\cite{litref29} develops a method to derive unit tests automatically by analyzing human-written specifications for preconditions and post-conditions.
\cite{litref30} discusses orchestrating domain-specific test languages with a BDD approach in the context of airplane systems testing.
\cite{litref31} examines behavior-driven requirements engineering in agile product line engineering, focusing on establishing and managing agile product lines.
\cite{litref32} integrates BDD with HW/SW Co-design in a renewable energy project, describing behavior as executable user stories.
\cite{litref33} presents BDD methodology and Cucumber framework for automating regression testing of Android apps, particularly testing the broadcast mechanism in Catrobat.
\cite{litref34} discusses re-engineering legacy systems with a transaction model, focusing on updating older software systems.
\cite{litref35} proposes extending BDD for hardware design, focusing on verification and increasing productivity in hardware testing.
\cite{litref36} models Gherkin scenarios using UML, aiming to enhance scenario definition and visualization in BDD.
\cite{litref37} evaluates BDD in software engineering, suggesting improvements to the process based on a case study.
\cite{litref38} reports on adopting BDD for an IoT system, using problem frames to define specifications at different system levels.
\cite{litref39} presents a framework for intuitive software test system design, focusing on acceptance tests and status reporting for business stakeholders.\cite{litref40} specifies software using DEMO transaction patterns and BDD, integrating them into initial BDD scenarios for enterprise information system development.\cite{litref41} proposes using BDD for acceptance testing in microservices architecture, building test scenarios for API REST services to match end-user needs. The paper focuses on a payroll system case study, assessing the Behave BDD tool for automating acceptance tests and highlighting its effectiveness in ensuring that the system meets user requirements.

\cite{litref42} explores the integration of Test-Driven Development (TDD) and Behavior-Driven Development (BDD) in GUI testing. The thesis proposes a novel approach using GUI prototypes for early testing, utilizing tools like JustInMind for capturing and replaying test cases. This method aims to enhance software quality from the design phase and align developer and customer perceptions, reducing the need for extensive test code writing.

\cite{litref43} discuss the challenges and opportunities in low-code testing in their paper. They analyze testing components of commercial Low-Code Development Platforms (LCDP) and propose a feature list for low-code testing, addressing the role of the citizen developer, the need for high-level test automation, and cloud testing.

\cite{litref44} assesses the re-usability in automated acceptance tests. It explores how automated acceptance tests, closely related to software requirements, can be reused by identifying characteristics through systematic literature reviews and case studies. The thesis proposes methods to measure the reuse potential of BDD test cases and a method to calculate cost avoidance through reuse, finding that automated acceptance tests are indeed reusable and their reuse potential can be quantified.

The study by \cite{litref45} is particularly relevant and the closest to the work in terms of other literature. It presents a reusable automated acceptance testing architecture for microservices in Behavior-Driven Development. While focusing on re-usability, their work revolves around using reusable step definitions, but this needs new step definitions to be written in a programming language. This paper, in contrast, aims to simplify and combine step definitions, making both the scenarios and the step definitions more accessible and broadly applicable. The paper introduces calling scenarios as a function, mapped to a step definition, following are the two main step definitions for the scope of this paper.

Though a short part of the literature review is relevant to this work, it does lay the foundation for presenting the work better. Though this work still builds on the insights and challenges identified in these studies, it ventures into the practical application of the presented enhanced approach to automated acceptance testing. 

\section{Implementation}
In this section the paper dives into the practicalities of developing and applying our proposed methodologies, demonstrating their effectiveness and efficiency in real-world software development scenarios.
\subsection{Reusable Scenario Step Definitions}
\begin{itemize}
    \item \textbf{I Call Feature File FolderFilesEnumeration(Objects, 1 for each):}
    \begin{itemize}
        \item Designed for scenarios requiring access to feature files from multiple folders.
        \item Useful for directories within the parent "features" directory, especially those starting with "reusable."
        \item Involves creating a copy of each step definition when a change is detected in folder structure of reusable features.
        \item Allows dynamic access to a range of feature files based on folder naming conventions.
        \item Alternative raw string usage: I call feature file "folderName" "fileNameInFolder".
    \end{itemize}

    \item \textbf{I Call Feature File FixedFolderFilesEnumeration}
    \begin{itemize}
        \item A simpler and more straightforward step definition.
        \item Relies on a fixed directory path specified within a configuration file.
        \item Focuses on facilitating the calling of scenarios from a predetermined location.
        \item Enhances ease of use and efficiency for routine or standard test scenarios.
        \item Alternative raw string usage: I call feature file "fileName".
    \end{itemize}
\end{itemize}

The second step definition is a subset of the first one. It offers a simplified approach by relying on a fixed directory path specified in the configuration file. This subset approach still maintains the versatility of the first definition but streamlines the process for more routine scenarios.

These step definitions are integral to an approach designed to enhance the auto-complete functionality in development tools like VSCode. The first step definition offers flexibility and access to multiple folders, while the second provides simplicity and speed through its reliance on a fixed directory. The implementation includes code that dynamically updates the step definitions in the Java file, ensuring seamless integration and facilitating more effective test case generation and execution.

\subsection{Definition of Dynamic ENUMs for Step Auto-Complete in VSCode}
The dynamic ENUM generation process is designed to facilitate step autocompletion in Visual Studio Code, particularly for Cucumber BDD scenarios. The process targets feature files located in directories named ``reusable'' or ``common'' within the ``/features/'' folder. The algorithm generates an ENUM class for each such directory, with ENUM values representing the concatenation of folder and file names.

\begin{algorithm}
\caption{Dynamic ENUM Generation}
\begin{algorithmic}[1]
\State $featuresPath \gets ``./features/''$
\State $targetDirs \gets [``reusable'', ``common'']$
\Procedure{GenerateEnums}{}
    \State $directories \gets$ list all subdirectories in $featuresPath$
    \For{each $dir$ in $directories$}
        \For{each $targetDir$ in $targetDirs$}
            \If{$dir$ name contains $targetDir$}
                \State \Call{GenerateEnumForFolder}{$dir$}
            \EndIf
        \EndFor
    \EndFor
\EndProcedure
\Procedure{GenerateEnumForFolder}{$folder$}
    \State $enumName \gets$ convert $folder$ name to Enum format
    \State $featureFiles \gets$ list all .feature files in $folder$
    \State Print ``public enum $enumName$ \{''
    \For{each $file$ in $featureFiles$}
        \State $enumConstant \gets$ uppercase($folder\_name$ + ``\_'' + $file$)
        \State Print $enumConstant$
    \EndFor
    \State Print ``\};''
\EndProcedure
\end{algorithmic}
\end{algorithm}

This Algorithm 1. outlines the steps for dynamically generating ENUM classes for Java, designed to work with the Cucumber Full Language Support extension [\cite{krechik2017cucumber}] in Visual Studio Code. The generated ENUMs will facilitate the autocompletion of Gherkin steps by mapping the folder and file structure within the specified feature file directories.

\begin{algorithm}
\caption{Generate Step Definitions}
\begin{algorithmic}[1]
\Procedure{GenerateStepDefinitions}{}
\State $featuresPath \gets ./features/''$ 
\State $targetDirs \gets [reusable'', common'']$ 
\State $directories \gets$ list all subdirectories in $featuresPath$ 
\For{each $dir$ in $directories$} 
\For{each $targetDir$ in $targetDirs$} 
\If{$dir$ name contains $targetDir$} 
\State $enumName \gets$ convert $dir$ name to Enum format 
\State $featureFiles \gets$ list all .feature files in $dir$ 
\For{each $file$ in $featureFiles$}
\State $enumConstant \gets$ uppercase($dir + '\_' + $file)
\EndFor
\State \Call{CreateStepDefinition}{$enumName$, $enumConstant$}
\EndIf
\EndFor
\EndFor
\EndProcedure

\Procedure{CreateStepDefinition}{$enumName$, $enumConstant$}
\State Print annotation @And("I call feature file $enumConstant$")
\State Print void callFeature + $enumName$ + ObjectName($enumName$ enumObject)
\State Print '\{'
\State \Call{CallReusableScenario}{$enumName$, $enumConstant$} using Cucumber static object
\State Print '\};'
\EndProcedure
\end{algorithmic}
\end{algorithm}

This Algorithm 2 iterates through the directories and their respective feature files, generating a step definition for each enum constant. The CreateStepDefinition procedure dynamically prints the step definition method with the necessary logic placeholder for calling the associated feature file. This approach will create a comprehensive set of step definitions corresponding to the enumerated feature files.
\\

\subsection{Gherkin Step Auto-Complete Functionality in VSCode}
The \cite{krechik2017cucumber} "Cucumber Full Language Support" extension for VSCode significantly enhances the development experience for those working with Cucumber (Gherkin) language, a popular tool for Behavior-Driven Development (BDD). This section of your paper aims to detail the usage, integration, and potential future applications of this VS Code extension.
\subsubsection{Installation and Setup}
Open Your Application in VS Code: Start by opening your application project in VS Code where you intend to use Cucumber for BDD.

Install the Extension: Search for the 'cucumberauto-complete' extension in the VS Code 
Extensions Marketplace and install it. This extension is designed to add comprehensive language support for Cucumber.

In your project's root directory, create a .vscode folder and a settings.json file within it (if they don't already exist). You can do this manually or by using the command \textbf{mkdir .vscode \&\& touch .vscode/settings.json} in your terminal.

Add Extension Settings to settings.json: Customize your Cucumber environment by adding necessary settings to the settings.json file. These settings can include paths to your feature files, regular expressions for step definitions, and language configurations. An example of how a .vscode/settings.json for the extension in java:
\begin{lstlisting}[style=java, caption=Settings for the VSCode  step definition auto-complete plugin.,label={lst:vscodesettings}]
{
    "cucumberauto-complete.stepsInvariants": true,
    "cucumberauto-complete.steps": [
        "**/*.java"
    ],
    "cucumberauto-complete.syncfeatures": "**/*.feature",
    "cucumberauto-complete.customParameters": [
        { 
            "parameter": "{string}",
            "value": "\".*\""
        },
        {
            "parameter": "{int}",
            "value": ".*"
        },
    ],
    "cucumberauto-complete.skipDocStringsFormat": true,
    "cucumberauto-complete.smartSnippets": true,
    "cucumberauto-complete.onTypeFormat": true,
    "editor.quickSuggestions": {
    "comments": true,
    "strings": true,
    "other": true
    },
    "cucumberauto-complete.gherkinDefinitionPart": "(Given|When|Then|But|And)\\(",
}
\end{lstlisting}
The detailed description of attributes from Listing~\ref{lst:vscodesettings} is present in \cite{krechik2017cucumber}.

Reload the Application: To apply the changes and activate the extension, reload your application in VS Code.
Utilizing Extension Features:

Syntax Highlighting and Snippets: The extension provides syntax highlighting for better readability and pre-defined code snippets for common Gherkin expressions.

Auto-Parsing and Autocompletion: Steps defined in your feature files are automatically parsed and made available for autocompletion, streamlining the coding process.

Validation and Definitions Support: The extension provides on-type validation for all steps and supports definitions, ensuring that your Gherkin syntax is both accurate and efficient.

Document Formatting: Includes formatting support for Gherkin documents, such as tables, making your feature files more organized and readable.

Multi-Language Support: It supports various programming languages (like JavaScript, TypeScript, Ruby, Kotlin, and Java) and multiple human languages, enhancing its utility in diverse development environments.

\textbf{Future Use Cases and Potential Enhancements}

Advanced AI-Driven Autocompletion: Future versions could incorporate AI to provide more intelligent and context-aware autocompletion suggestions based on the specific usage patterns and domain language of the project.

Integrated Testing and Reporting: Integration with testing frameworks and reporting tools directly within VS Code could provide real-time feedback and detailed test reports, enhancing the TDD/BDD workflow.

Collaboration Features: Future enhancements might include real-time collaboration tools for teams working on BDD, including shared editing and live commenting within feature files.

Enhanced Refactoring Tools: Enhanced tools for refactoring steps and features, including automatic updates of associated step definitions and feature files when changes are made, could significantly streamline the development process.

Customizable Workflow Automation: Introducing features that allow teams to automate repetitive tasks in the BDD process, customized to their workflow, can further increase productivity.

Integration with Version Control Systems: Better integration with version control systems like Git for managing changes in feature files and associated step definitions could be a valuable addition.

Learning and Suggestion Mechanism: The extension could learn from user corrections and preferences over time, offering more personalized and accurate suggestions and corrections.

In summary, the Cucumber Full Language Support extension for VS Code is a powerful tool that enhances the BDD process by providing comprehensive language support, autocompletion, validation, and formatting for Gherkin. As BDD continues to grow in popularity, extensions like this will become increasingly vital for efficient and effective software development workflows.

\subsection{Running Reusable Steps in Jenkins}

For local test execution in VSCode, especially to utilize JUnit test pane in VSCode, the following code structure is typically employed:

\begin{lstlisting}[style=java, caption=Sample Java code for JUnitRunner.java file mentioned later in the paper.]
@RunWith(Cucumber.class)
@CucumberOptions(features = {
    "PathToFeatureFiles" }, glue = {
    "PathToStepDefinitionsJavaDirectory" }, plugin = { "html:target/cucumber-html-report",
    "json:target/cucumber.json",
    "pretty:target/cucumber-pretty.txt",
    "junit:target/test-results.xml" }, dryRun = false, tags = { "@ScenarioTag" })
public class TestRunner {

    @BeforeClass
    public static void setup() throws Exception {
        // Method contents omitted
    }

    @AfterClass
    public static void teardown() {
        // Method contents omitted
    }
}
\end{lstlisting}

It is important to note that each value within the @Annotation must be a fixed string for the Java Virtual Machine (JVM) to execute it correctly.

In a continuous integration environment like Jenkins, the test execution is configured as a Maven project and is executed generally using the following command:

\begin{lstlisting}[style=java, caption=Command to run Maven test on Jenkins(cmd prompt command)]
test -Dcucumber.options="PathToFeatureFiles --tags @ScenarioTag" -Denv=\$env
\end{lstlisting}

In this setup, the @ScenarioTag refers to a scenario or multiple scenarios located within the PathToFeatureFiles, particularly those containing reusable steps.

A challenge arises when there is a need to temporarily set global cucumber.options, which is not permissible in Cucumber. This issue becomes evident particularly when calling the scenario function again, resulting in the re-execution of the @ScenarioTag step from the beginning.

To address this, an alternative solution that does not rely on global cucumber.options is required. Considering that each Jenkins job has a specific purpose and clones code from the master branch in a Git repository, the use of @CucumberOptions was dynamically updated. This approach involves replacing a placeholder with a variable specified in the Jenkins job. Consequently, the command for test execution can be run without explicitly setting cucumber.options that set the global values for Cucumber variables.
In a Jenkins continuous integration setup, a Bash script is commonly used to prepare the testing environment before running the Maven test command. The following Bash commands demonstrate how scripts are executed within Jenkins:

\begin{lstlisting}[style=bash, caption=Bash code to update JUnitRunner.java file annotation.]
chmod +x jenkins.sh
./jenkins.sh "$PathToFeatureFile" "$ScenarioTag"
\end{lstlisting}

The "jenkins.sh" script is designed to modify the placeholders in the @CucumberOptions annotation in the main JUnit runner file. The contents of this script are tailored to suit the specific needs of the testing environment and may include various shell commands and scripting logic. The code for 'run.sh' file is as follows:

\begin{lstlisting}[style=bash, caption=Contents of the "jenkins.sh" file.]
# Get input parameters
featurePath=$1
tags=$2

TestFilePath=JUnitRunner.java

sed -i "s|\"$PathToFeatureFile \"|\"${featurePath}\"|g" $TestFilePath
sed -i "s|\"@ScenarioTag\"|\"${tags}\"|g" $TestFilePath

\end{lstlisting}

This approach allows for dynamic configuration of test parameters, facilitating flexibility and adaptability in automated testing scenarios.

\section{Usecase}
\subsection{E-commerce Website}
In the dynamic environment of e-commerce, efficient and robust testing methodologies are essential for ensuring seamless user experiences, especially in critical functionalities like item search and checkout processes. This section delves into the practical application of reusable scenarios in the context of an e-commerce website, demonstrating how the enhanced Behavioral Driven Development (BDD) methodology can streamline and optimize the testing process.

The primary advantage of using reusable scenarios in an e-commerce setting is the ability to define a set of common steps that can be applied across multiple test cases. This approach not only reduces redundancy but also ensures consistency and ease of maintenance. Particularly for e-commerce platforms, where numerous items share similar pathways through the site's functionality, the ability to reuse test scenarios becomes invaluable.

Consider the following example of a reusable step in Gherkin syntax:

\begin{lstlisting}[style=gherkin, caption={Use of reusable scenario in Gherkin syntax for testing a Feature in various ways for an e-commerce platform. This example illustrates the usage of a common/reusable scenario for different items with reusable/unreusable steps for each.}]
@FeatureX @FeatureX_Test1
Scenario: Testing {feature}(common) for {first} item
    Given I start the testcase
    When I go to the item search page
    Then I search for {first} item
    And I click on {first} item
    And I call feature file enumConstant [common]
    And I do any step left specific to {first} item...
    Then Test Ends

@FeatureX @FeatureX_TestN
Scenario: Testing {feature}(common) for {Nth} item
    Given I start the testcase Nth
    When I go to the item search page
    Then I search for {Nth} item
    And I click on {Nth} item
    And I call feature file enumConstant [common]
    And I do any step left specific to {Nth} item...
    Then Test Ends
\end{lstlisting}

In this example, the `{feature}` placeholder represents a common functionality being tested, such as the checkout process. The `{first} item` and `{Nth} item` are placeholders for specific items on the e-commerce platform. The `enumConstant [common]` denotes the reusable scenario, which includes a series of steps common to all items being tested. This structure allows for the testing of multiple items with minimal scenario duplication, enhancing efficiency and reducing the potential for errors in test script creation.

In summary, the use of reusable scenarios in an e-commerce website context exemplifies the efficiency and effectiveness of the enhanced BDD methodology. This approach not only streamlines the testing process but also contributes to maintaining high standards of quality and reliability in a highly competitive digital marketplace.

\section{Primary Limitation of the Approach: Complexity Management}

While the proposed approach offers substantial benefits, it does introduce a level of complexity that requires careful management:

\subsection{Complexity in Integration and Maintenance}
The integration of various components such as Cucumber, Jenkins, and VSCode’s auto-complete functionality, requires a comprehensive understanding of these systems. This complexity, however, can be effectively managed through proper documentation and team training.

\subsection{Balancing Sophistication with Usability}
Ensuring that the advanced capabilities of the approach do not compromise usability is crucial. By offering detailed guides and support, teams can leverage the full potential of the methodology without being overwhelmed.

\subsection{Facilitating Smooth Transition for Teams}
For teams transitioning from traditional BDD practices, the shift to this more advanced methodology can be made smoother through structured training programs and gradual implementation strategies.

\subsection{Maintaining Scalability and Clarity}
As the framework expands, maintaining its scalability and clarity will be essential. Regular reviews and updates of the test suites can help manage this complexity, ensuring the framework remains effective for large-scale projects.

\section{Conclusion}
This study presents a significant advancement in the field of Behavioral Driven Development (BDD), particularly in the context of scenario re-usability within the Cucumber framework. The introduction of methods allowing for the execution of reusable scenarios within a single-threaded process marks a notable improvement in the efficiency and modularity of BDD test suites. The implementation of these methodologies in Java, with potential applicability in other programming languages, demonstrates versatility and broad scope.

The integration of VSCode’s step definition auto-complete functionality emerges as a key factor in simplifying the scenario writing process. This feature proves especially beneficial for new QA test writers, facilitating an easier and more efficient workflow. The study's exploration into the application of these methodologies within Jenkins, a prominent continuous integration and delivery platform, showcases a novel approach to adapting Cucumber’s functionalities to various testing environments.

Empirical research and practical applications indicate substantial improvements in test development speed and execution efficiency. These enhancements are crucial for large-scale and complex software projects, where conventional BDD practices may be less effective. Overall, the integration of these innovative methods with traditional BDD practices points towards a future of more effective, efficient, and sustainable test automation strategies. The emphasis on scenario re-usability within the Cucumber framework is particularly promising, suggesting new directions for research and development in the field.

\section{Conflict of Interest and Permissions}

\textbf{Conflict of Interest:} The foundational ideas for this research originated from discussions with Technology Management at Xpressdocs. While the initial concept was developed in a professional setting, some ideation and development for this paper occurred independently. The framework, with certain modifications, is employed at Xpressdocs for Cucumber BDD testing. The author have ensured that no proprietary or confidential information from Xpressdocs is included in this paper. This measure was taken to avoid any potential conflict of interest and to maintain the integrity of the research. Despite the professional association with Xpressdocs, there has been no commercial or financial influence that affected the outcomes of this paper.

\textbf{Permissions and Ethical Compliance:} The author confirms that all necessary permissions were obtained from company officials at Xpressdocs prior to the publication of this paper. This was done to ensure adherence to and respect for proprietary information. The research presented here adheres to the ethical guidelines and policies set forth by my institution and aligns with the agreements made with Xpressdocs. Obtaining these permissions highlights my commitment to ethical research practices and ensures transparency in the dissemination of my findings. The author affirms that this research is conducted and shared with the utmost respect for intellectual property and organizational confidentiality.









\bibliographystyle{unsrtnat}
\bibliography{paper}  






\end{document}